Improved test-retest reliability of $R_2^*$ and susceptibility quantification using multi-shot multi-echo 3D EPI

Yujia Huang[1], Lin Chen[2,3], Xu Li[2,3] and Jiaen Liu[1,4,*]

1 Advanced Imaging Research Center, UT Southwestern Medical Center, Dallas, TX, USA

2 F.M. Kirby Research Center for Functional Brain Imaging, Kennedy Krieger Institute, Baltimore, MD, USA

3 Department of Radiology and Radiological Sciences, Johns Hopkins University, Baltimore, MD, USA

4 Department of Radiology, UT Southwestern Medical Center, Dallas, TX, USA

**Abstract**

This study aimed to evaluate the potential of 3D echo-planar imaging (EPI) for improving the reliability of $T_2^*$-weighted ($T_2^*$w) data and quantification of $R_2^*$ decay rate and susceptibility ($\chi$) compared to conventional gradient echo (GRE)-based acquisition. Eight healthy subjects in a wide age range were recruited. Each subject received repeated scans for both GRE and EPI acquisitions with an isotropic 1 mm resolution at 3 T. Maps of $R_2^*$ and $\chi$ were quantified and compared using their inter-scan difference to evaluate the test-retest reliability. Inter-protocol differences of $R_2^*$ and $\chi$ between GRE and EPI were also measured voxel by voxel and in selected ROIs to test the consistency between the two acquisition methods. The quantifications of $R_2^*$ and $\chi$ using EPI protocols showed increased test-retest reliability with higher EPI factors up to 5 as performed in the experiment and were consistent with those based on GRE. This result suggested multi-shot multi-echo 3D EPI can be a useful alternative acquisition method for $T_2^*$w MRI and quantification of $R_2^*$ and $\chi$ with reduced scan time, improved test-retest reliability and similar accuracy compared to commonly used 3D GRE.

**Introduction**

Magnetic susceptibility ($\chi$) of brain tissue provides useful information associated with pathological changes such as activity of immune cells (Connor et al., 1992; Ward et al., 2014), demyelination (Kim et al., 2023), iron accumulation (Langkammer et al., 2012; Murakami et al., 2015; Chen et al., 2021), etc. It can be inferred noninvasively using magnetic resonance imaging (MRI) methods such as the effective transverse relaxation rate $R_2^*$ mapping and quantitative susceptibility mapping (QSM) (Haacke et al., 2009; Sati et al., 2013; Wang & Liu, 2015) using multi-echo $T_2^*$-weighted ($T_2^*$w) data. Importantly, the susceptibility contrast increases with the magnetic field ($B_0$) strength and can be exploited to develop submillimeter high spatial resolution imaging with sufficient contrast to noise ratio (CNR) during clinically practical scan time at 7 T (Duyn et al., 2007).

*Correspondence: Jiaen Liu, Advanced Imaging Research Center and Department of Radiology, UT Southwestern Medical Center, 2201 Inwood Rd. Dallas, TX, 75390 U.S.A. Email: Jiaen.Liu@UTSouthwestern.edu

Conventionally, $T_2^*$w data has been acquired using long TE gradient echo (GRE) sequence. It is time consuming because the minimum TR is limited by the optimal TE which is equal to the $T_2^*$ in the order of 30 ms for brain tissue at 7T and even longer at 3T, and only one k-space line is acquired in each TR. For example, a whole brain scan with voxel volume of 1 mm$^3$ takes between 5 to 10 min (Wang & Liu, 2015). This limits the application of $T_2^*$w MRI at even higher resolution. In addition, data acquired in a longer period is more sensitive to motion and $B_0$ fluctuation the latter of which can be caused by motion and other physiological sources (Van de Moortele et al., 2002; Liu et al., 2018).

The main hypothesis in this study was that improved test-retest reliability of $R_2^*$ and $\chi$ quantification can be achieved with a fast $T_2^*$w MRI acquisition method. One alternative for multi-echo $T_2^*$w data acquisition is the multi-shot 3D echo-planar imaging (EPI). The scan time can be reduced by acquiring multiple k-space lines for each echo in one TR, which is the so-called EPI factor. Unlike acceleration using parallel imaging, 3D EPI does not introduce the g-factor penalty (Robson et al., 2008) which is more deteriorating with high parallel imaging factors. Although 3D EPI has been applied to QSM in previous works showing similar results as the 3D GRE method (Langkammer et al., 2015; Wicaksono et al., 2021), its test-retest reliability compared to 3D GRE has not been quantitatively evaluated.

In this study, we tested the hypothesis in healthy subjects of different ages at 3 T. The test-retest reliability of the $T_2^*$w image magnitude, $R_2^*$ and $\chi$ was evaluated and compared among multi-echo 3D EPI with various EPI factors and 3D GRE. The similarity of $R_2^*$ and $\chi$ between EPI and GRE was also evaluated. The contributions of thermal noise, which was more significant in the shorter EPI sequence, and physiological noise were analyzed in the observed test-retest reliability data in the two methods.

**Methods**

MRI experiment

Experiments were performed on a 3 T MRI scanner (Prisma, Siemens, Erlangen, Germany) using the vendor-provided 64-channel head-neck receive array RF coil. Eight healthy subjects were recruited with signed consent (18 to 68 [46±20] years old, 4 males) under a research protocol approved by the Institutional Review Board. The imaging protocol included $T_2^*$w multi-echo 3D GRE and multi-echo 3D EPI sequences (Table 1 and Fig. 1). The major difference between GRE and EPI protocols was the EPI factor, which defines the number of acquired k-space lines per echo. All protocols were accelerated based on the controlled aliasing in parallel imaging (CAIPI) technique (Breuer et al., 2006) using a 2×2 acceleration factor in the first and second phase encoding directions as shown in Fig. 1. The acquisition bandwidth was chosen such that the total data acquisition time for each echo was matched as much as possible between EPI and GRE (Table 1). This is allowed by simply repeating the EPI acquisitions by the number of the corresponding EPI factors, giving matched signal to noise ratio (SNR). For EPI, these repetitions formed one scan. Scans were performed twice for the test-retest reliability analysis. The order of the protocols was randomized in different subjects to avoid bias due to subject fatigue.

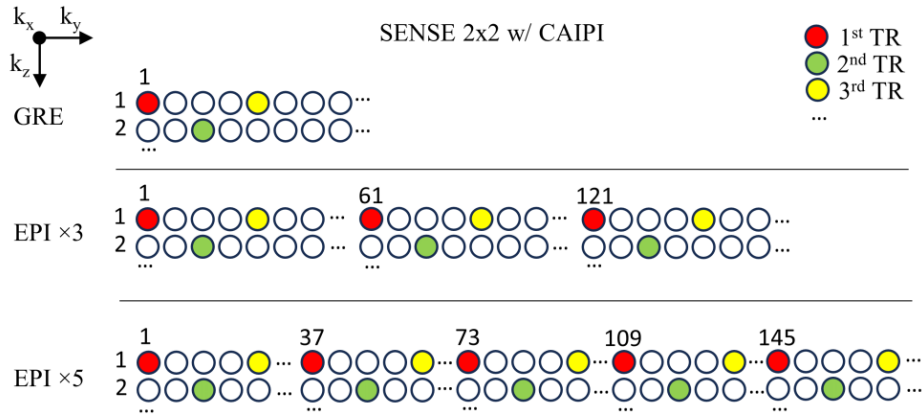

**Fig. 1** Acquisition pattern of multi-echo 3D GRE and multi-echo 3D EPI for a given acceleration factor of 2×2 with CAIPI in $k_y$-$k_z$ plane. Solid color dots represent the first three shots of the acquired k-space lines.

In all scans, 3D EPI navigators were obtained between the RF excitation and the regular data acquisition using a segmented 3D EPI trajectory for motion and $B_0$ correction, similar to the previous work (Liu et al., 2020; van Gelderen et al., 2023). Here, the navigator was acquired using matrix size of 40×32×12, spatial resolution of 6×5.6×15 mm$^3$ and acceleration factor of 4×2 with CAIPI. Eight k-space lines of the navigator were acquired during each TR in less than 5 ms, and it took 6 TRs to acquire one navigator image.

A $T_1$-weighted magnetization prepared rapid acquisition gradient echoes (MPRAGE) sequence was acquired with isotropic 1 mm resolution for tissue segmentation and group analysis.

To reconstruct the accelerated GRE and EPI data, a reference scan was performed using a 2D GRE sequence with flip angle of 40°, TE of 2.5 ms, TR of 270 ms, in-plane resolution of 4 mm, slice thickness of 5 mm, FOV of 240×180×200 mm$^3$ and scan time of 12 s.

|  | 3D GRE | 3D EPI (×3) | 3D EPI (×5) |
|---|---|---|---|
| Isotropic resolution (mm) | 1 | 1 | 1 |
| Field of view (mm) | 240×180×144 | 240×180×144 | 240×180×144 |
| SENSE factor | 2×2+CAIPI | 2×2+CAIPI | 2×2+CAIPI |
| TE1/TE4/ΔTE (ms) | 14/47/11 | 13/49/12 | 13/52/13 |
| TR (ms) | 60 | 64 | 66 |
| Flip angle (°) | 20 | 20 | 20 |
| Readout bandwidth (Hz/pixel) | 125 | 372 | 631 |
| Readout train length (ms) | 8 | 10 | 10 |
| EPI factor | 1 | 3 | 5 |
| Scan time per repetition | 7'54" | 2'49" | 1'42" |
| Repetitions | 1 | 3 | 5 |
| Data acquisition time per scan (s) | 254 | 255 | 251 |
| Scans | 2 | 2 | 2 |

Note: CAIPI stands for controlled aliasing in parallel imaging (Breuer et al., 2006). The GRE protocol was based on a unipolar readout gradient waveform.

**Table 1** Parameters of $T_2^*$-weighted GRE and EPI protocols with different EPI factors.

Image reconstruction

The $T_2^*$w GRE and EPI images were reconstructed with custom MATLAB (Mathworks, Natick, MA, USA) software which used the navigator information to perform motion and $B_0$ correction. Details about the reconstruction can be found in the previous publication (van Gelderen et al., 2023). Briefly speaking, the reconstruction model considered the effect of rigid body motion and spatially linear $B_0$ change for artifact correction, and the receive $B_1$ sensitivity maps for parallel imaging reconstruction. The solution was built on the non-uniform fast Fourier transform algorithm (Fessler & Sutton, 2003) by considering the effect of motion and linear $B_0$ changes on the actual sampling coordinate and signal phase in the k-space. Originally developed for GRE reconstruction, it was expanded for EPI reconstruction by applying the measurement of motion and linear $B_0$ changes to multiple k-space lines acquired in the same TR. For EPI data, a phase correction compensating for the alternating polarity of the readout gradient was performed using calibration data with no phase encoding, i.e., $k_y=k_z=0$, acquired at the beginning of each EPI scan.

All images were reconstructed in the corrected and uncorrected modes. The corrected mode included modeling the effect of motion and linear $B_0$ changes while the uncorrected mode only accounted for the global average $B_0$ change. Motion was estimated based on the navigator magnitude image by minimizing the mean square magnitude difference using an iterative multi-resolution approach (Thevenaz et al., 1998). The $B_0$ changes were estimated using the aligned and unwrapped navigator phase changes over time with the nominal navigator TE of 3.9 ms. Evaluation of the navigator accuracy for motion and field measurement can be found in the previous work (Liu et al., 2020; van Gelderen et al., 2023).

Image processing and data analysis

The decay rate $R_2^*$ was calculated by fitting a mono-exponential function to the multi-echo magnitude data based on the least square nonlinear fitting in MATLAB (lsqcurvefit). Quantitative susceptibility mapping (QSM) was carried out using the JHU/KKI QSM toolbox v3.3, (https://github.com/xuli99/JHUKKI_QSM_Toolbox) (Bao et al., 2016; Li et al., 2019; van Bergen et al., 2016). The entire process included the following steps: best path-based phase unwrapping (Abdul-Rahman et al., 2005), brain masking using FSL BET (Smith, 2002), combined LBV (Zhou et al., 2014) and VSHARP (Wu et al., 2012) for background field removal, echo combination using weighted echo averaging (Wu et al., 2012) and dipole inversion using a modified structural feature collaborative reconstruction method (Bao et al., 2016) using the L2 norm of a nonlinear data fidelity cost function (Milovic et al., 2018). The tissue reference value for susceptibility quantification was the average of all the brain tissues included in the brain mask.

Test-retest reliability of the imaging data was quantified based on the normalized inter-scan absolute difference (NAD) of the magnitude images. NAD was calculated as the absolute difference between two magnitude images divided by their mean on a voxel-by-voxel basis. For EPI, additional averaging of the magnitude images across all repetitions within each scan was performed to retain equal SNR as the GRE magnitude data. Rigid-body image transformation was carried out to correct for inter-scan motion using the aforementioned multi-resolution coregistration approach (Thevenaz et al., 1998) followed by spline interpolation in the software of FSL (FLIRT tool, fsl.fmrib.ox.ac.uk). The magnitude data was chosen because the phase data for each scan could contain uncertain phase offset.

Test-retest reliability of the $R_2^*$, tissue frequency shift ($f$) and $\chi$ was quantified based on the absolute inter-scan difference $|\Delta R_2^*|$, $|\Delta f|$ and $|\Delta \chi|$. They were calculated based on the mean $R_2^*$, $f$ and $\chi$ with different number of averages over the first few repetitions of data within each EPI scan. Here, the mean was calculated over the $R_2^*$, $f$ and $\chi$ from individual repetitions rather than based on the averaged MRI data. This allowed us to evaluate the noise contribution to the test-retest reliability. It was expected that with fewer averages, the EPI result would be more dominated by the thermal noise effect but could still achieve satisfactory overall test-retest reliability (van der Zwaag et al., 2012). Inter-scan motion was corrected using the same coregistration parameters derived from the magnitude images as mentioned earlier.

Consistency of $R_2^*$ and $\chi$ between GRE and EPI protocols were evaluated voxel by voxel and in selected deep brain ROIs at the group level. For the voxelwise analysis, the $R_2^*$ and $\chi$ maps were first corrected for protocol-specific spatial distortion due to the $B_0$ inhomogeneity, with correction in the frequency encoding direction for the GRE data and the phase encoding direction for the EPI data in FSL (FUGUE tool) using the measured $B_0$ map. Note that the distortion correction was not applied in the test-retest analysis because the distortion was protocol specific. The distortion correction was followed by coregistration of the $R_2^*$ and $\chi$ maps based on the EPI magnitude data in reference to the GRE magnitude data. For ROI-based analysis, deep gray and white matter ROIs were selected from a previously developed QSM atlas in the MNI space (Li et al., 2019; Chen et al., 2021). This atlas was developed based on QSM images of 30 subjects (aged 71.2 ± 7.1 years, 21 females, nine males) in the MNI space. In the atlas, deep gray matter ROIs were defined based on the QSM contrast, and cortical grey matter and white matter ROIs were segmented based on coregistered $T_1$ MPRAGE image. Here, the selected ROIs included 5 deep grey nuclei (substantia nigra, red nucleus, putamen, globus pallidus and caudate nucleus) and 3 white matter regions (splenium of corpus callosum, posterior limbs of internal capsule and optic radiation).

Simulation of test-retest reliability

In order to analyze the effect of thermal noise in the test-retest reliability, Monte Carlo simulation was utilized by adding complex noise to one subject's complex GRE data. Because the noise level in the EPI data could be predicted from the GRE noise level with known readout bandwidth, the simulation was only performed based on the GRE data. The noise level was determined using the measured noise data without RF excitation, sensitivity maps and acceleration factors (Pruessmann et al., 1999). NAD, $|\Delta R_2^*|$ and $|\Delta f|$ were quantified in the simulation. Note that since $|\Delta \chi|$ could be specific to the regularization algorithm in the QSM dipole inversion, it was not evaluated in the simulation.

**Results**

The test-retest reliability result of the $T_2^*$w magnitude data is shown in Fig. 2 in one slice in the MNI space at the group level. It can be seen that the inverse of the NAD (NAD$^{-1}$) map increased with higher EPI factors, suggesting increased test-retest reliability. Note that the GRE protocol was essentially an equivalent EPI protocol with the EPI factor of 1. Increased test-retest reliability can also be appreciated in the corrected reconstruction compared to the uncorrected reconstruction in all protocols.

Besides the slice shown in Fig. 2, the NAD result including all voxels in all subjects was summarized and shown in Fig. 3. In this figure, the cumulative distribution function (CDF) described the fraction of voxels (y-axis) below a certain specific NAD value (x-axis). It can be seen that with higher EPI factors, a larger fraction of voxels was observed with lower NAD value, indicating increased test-retest reliability consistent with the result in Fig. 2. This plot also indicated the median NAD value for all cases corresponding to the CDF value of 0.5. Furthermore, the NAD increased at longer echo time due to lower signal amplitude and higher sensitivity to field fluctuation, and the result was improved in the corrected vs. the uncorrected reconstruction. It is also noticeable that the improvement with higher EPI factors was more significant in the uncorrected reconstruction data. As a reference, the CDF curve reflecting the thermal noise effect based on the Monte Carlo simulation was shown.

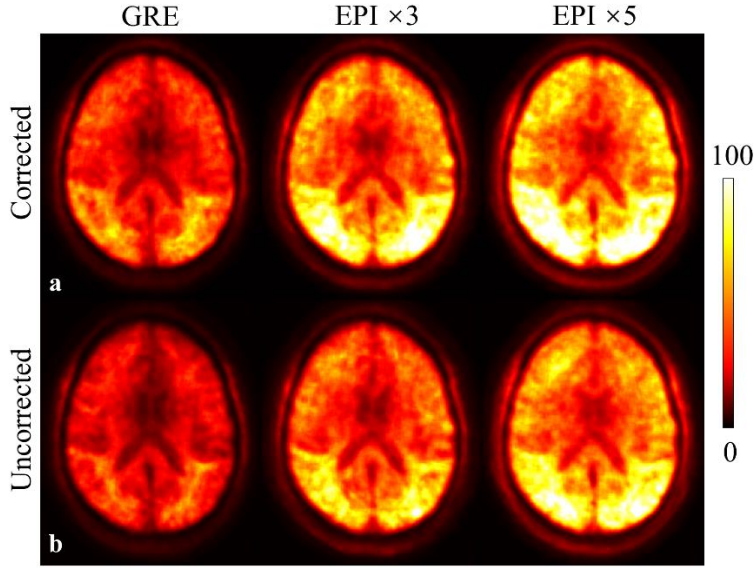

**Fig. 2** Inverse inter-scan normalized absolute difference (NAD$^{-1}$) maps of the magnitude data in the MNI space for the motion and B$_0$ corrected (a) and uncorrected (b) reconstruction. The result was derived from the average magnitude images across all echoes and across all EPI repetitions within each scan.

Improved test-retest reliability in the $R_2^*$, $f$ and $\chi$ quantification was also observed in the EPI protocols compared with GRE as shown in Fig. 4. Here, the voxelwise cumulative distribution of the inter-scan absolute difference of $|\Delta R_2^*|$, $|\Delta f|$ and $|\Delta \chi|$ were summarized for all subjects. It shows that with increased EPI factors, a larger portion of voxels exhibited reduced inter-scan differences. This trend can be seen in both corrected and uncorrected data. The median $|\Delta R_2^*|$ and $|\Delta f|$ caused by thermal noise were also shown in Fig. 4, suggesting the EPI test-retest reliability approaching the noise limit.

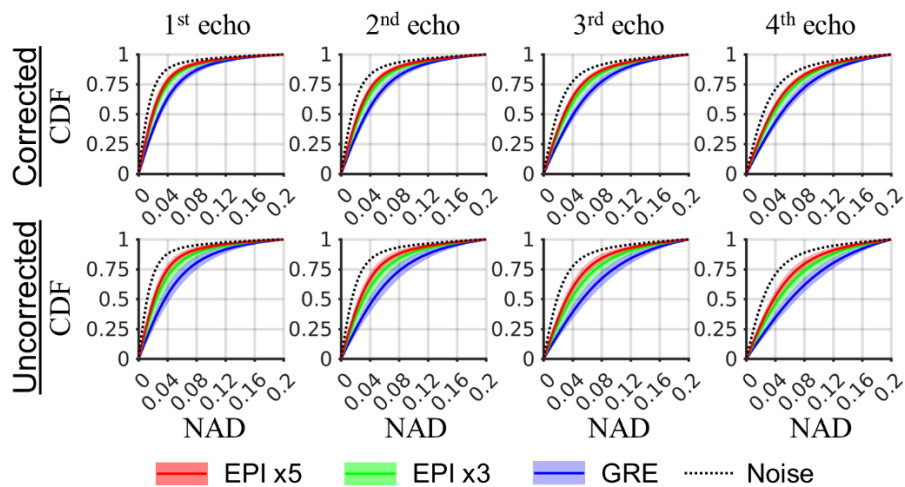

**Fig. 3** Cumulative density functions (CDF) of the inter-scan normalized absolute difference (NAD) result for individual echo magnitude images of all subjects. The equivalent NAD due to thermal noise (dotted line) was based on simulation. The shaded area marks standard deviation of the CDF over subjects.

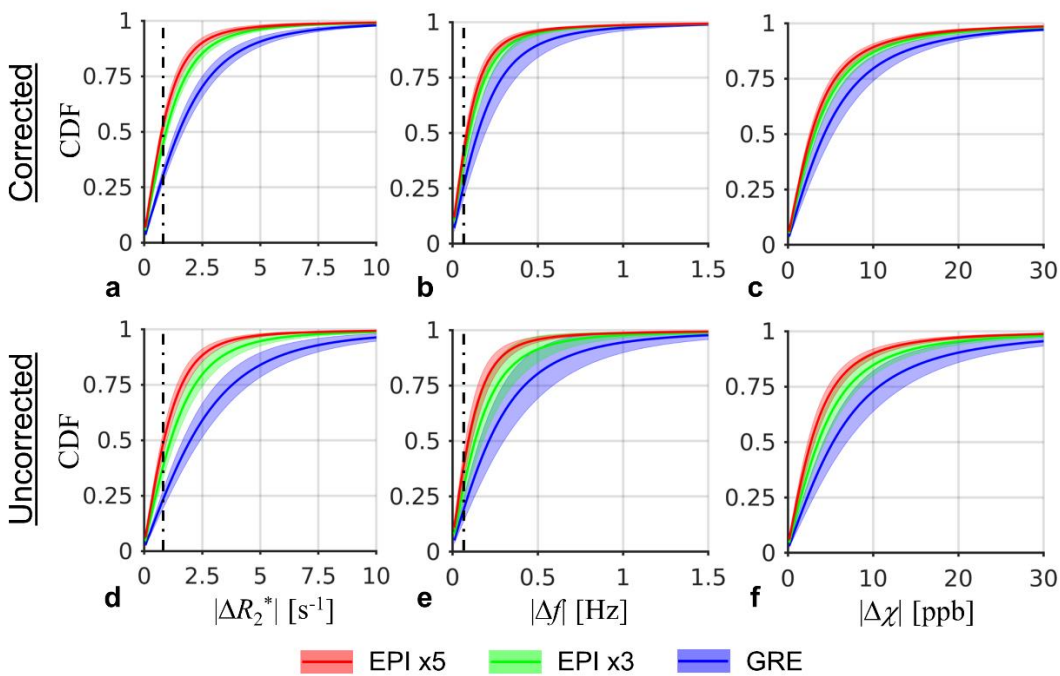

**Fig. 4** Cumulative density functions (CDF) of the inter-scan absolute differences of $R_2^*$, $f$ and $\chi$. Dash line represents the median difference due to noise based on simulation. The shaded area marks standard deviation of the CDF over subjects.

In addition to the test-retest reliability based on repetition-averaged data as described above, the result of median $|\Delta R_2^*|$ and $|\Delta\chi|$ with different averages was shown in comparison with that of GRE in Fig. 5.

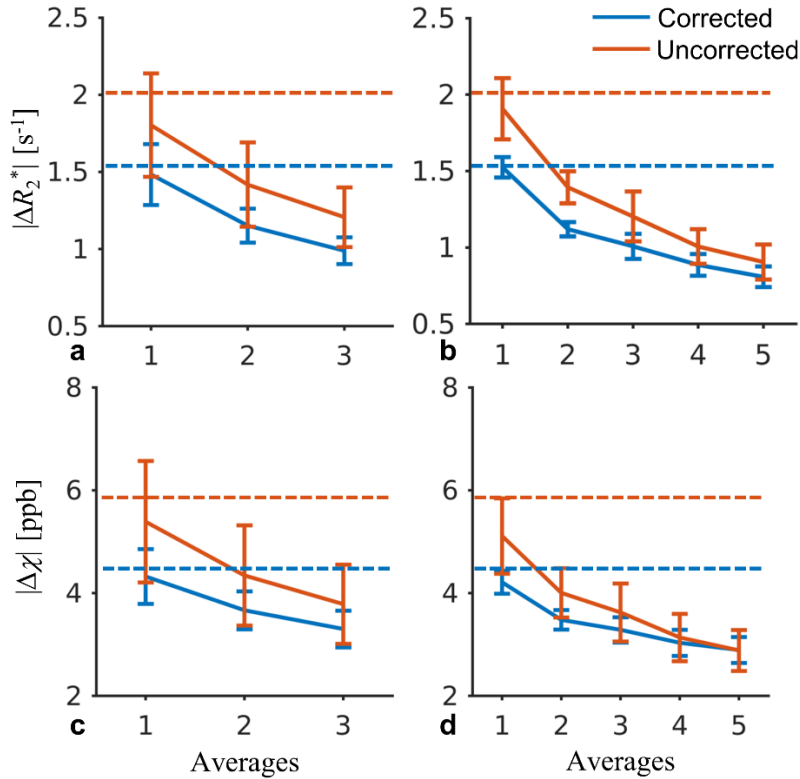

**Fig. 5** Median value of the inter-scan absolute differences of $R_2^*$ and $\chi$ as a function of the number of averages at the group level using multi-shot multi-echo 3D EPI. The group averaged GRE result is indicated by the dash lines. The error bars mark the standard deviation across subjects.

Averaging increased the test-retest reliability of $R_2^*$ and $\chi$ by improving the SNR. Interestingly, even without averaging, the $|\Delta R_2^*|$ and $|\Delta\chi|$ appeared to be similar to those of GRE, suggesting that the uncertainty of $R_2^*$ and $\chi$ in the GRE data was dominated by physiological noise.

Importantly, the result of $R_2^*$ and $\chi$ using the EPI protocols was consistent with the result based on the GRE sequence. In Fig. 6 A and C, representative $R_2^*$ and $\chi$ images of one subject are shown, demonstrating very similar results across all protocols. This can also be appreciated from the absolute difference maps in Fig. 6 B and D. More significant differences can be noticed in regions where the image signal was low (e.g., skull and veins). In the group level, the median voxelwise absolute $R_2^*$ difference was 1.7±0.3 s$^{-1}$ (EPI×5 vs. GRE) and 1.7±0.4 s$^{-1}$ (EPI×3 vs. GRE) in comparison to 1.4±0.3 s$^{-1}$ between repeated GRE scans (GRE vs. GRE), and the absolute $\chi$ difference was 5±1 ppb (EPI×5 vs. GRE) and 5±1 ppb (EPI×3 vs. GRE) in comparison to 4±1 ppb (GRE vs. GRE).

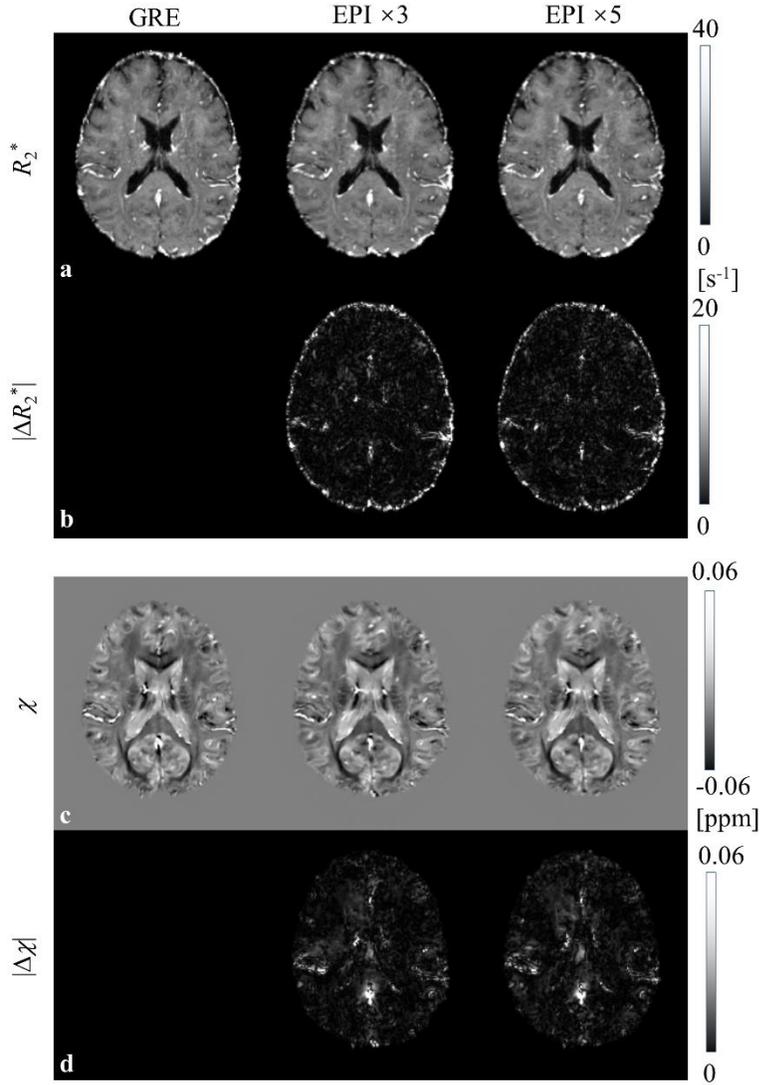

**Fig. 6** Representative images of $R_2^*$ (a) and $\chi$ (c) from 3D GRE and 3D EPI, and their absolute differences (b and d) relative to the GRE result from one subject. All images are acquired in a single repetition.

In several selected ROIs, the ROI-mean $R_2^*$ and $\chi$ were compared between the EPI and GRE data as shown in Fig. 7. The correlation coefficient (CC) and slope of the ROI-average $R_2^*$ across protocols were

0.9952 and 1.0002 (EPI×5 vs. GRE) and 0.9943 and 0.99837 (EPI×3 vs. GRE), respectively. The CC and

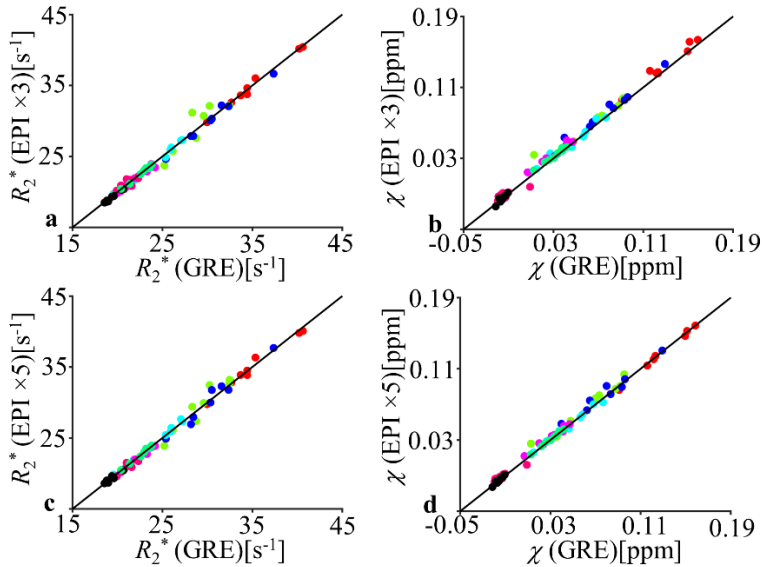

**Fig. 7** Comparison of ROI-based $R_2^*$ and $\chi$ between EPI and GRE data. Each data point represents one ROI average from one subject. The identity line is shown as the solid line. ROIs include 5 deep grey nuclei (substantia nigra [blue], red nucleus [green], putamen [cyan], globus pallidus [red] and caudate nucleus[magenta]) and 3 white matter regions (splenium of corpus callosum [pink], posterior limbs of internal capsule [spring green] and optic radiation [black]).

slope of the ROI-average $\chi$ were 0.9972 and 1.0107 (EPI×5 vs. GRE) and 0.9961 and 1.0481 (EPI×3 vs. GRE), respectively with $p<0.001$.

**Discussion**

In this study, the test-retest reliability of $T_2^*$w image, $R_2^*$ and $\chi$ were evaluated and compared between multi-shot multi-echo 3D EPI and GRE protocols at 3 T in 8 healthy subjects in a wide age range. The result showed higher reliability using the 3D EPI sequence with increasing EPI factors. In addition, the $R_2^*$ and $\chi$ maps were shown to be similar between the EPI and GRE data, suggesting that 3D EPI is a potentially useful alternative to GRE for clinical applications when reproducibility is often desired.

Motion and field correction plays a critical role for high resolution $T_2^*$w MRI (Tisdall et al., 2012; Gretsch et al., 2018; Liu et al., 2020; van Gelderen et al., 2023), achieving improved test-retest reliability as shown in the current result and a previous work (van Gelderen et al., 2023). Here, it was demonstrated that further improvement can be achieved by reducing the scan time, e.g., with increased EPI factors. As shown in Fig. 4, the inter-scan difference in the corrected EPI data with EPI factor of 5 approached the thermal noise floor. In Fig. 5, even without averaging, the EPI test-retest reliability performed similarly to the GRE method although the EPI SNR was much lower, suggesting the GRE data was more significantly affected by physiological noise. The current result agreed with a previous study, in which improved temporal SNR in 3D EPI time series with higher EPI factors was found and attributed to reduced physiological noise by analyzing human and phantom data (van der Zwaag et al., 2012).

It is worth mentioning that the EPI data in the second scan was aligned to the first scan with only one rigid-body transformation in the same way as the GRE data. This avoided potential bias due to the low-pass filtering effect in the transformation algorithm in the test-retest reliability result. It also suggests the test-retest reliability can be further improved by aligning the individual EPI images because motion parameters from such high-resolution images are expected to be more accurate in correcting for the inter-repetition motion. So far, the inter-repetition motion was only corrected during the image reconstruction based on the low-resolution navigator.

In the current GRE protocol, four echoes ranging from 13 to 52 ms were acquired (Table 1). The echo time range was in agreement with common GRE protocols (Wang & Liu, 2015) albeit a slightly lower number of echoes were obtained. This was due to our motivation to match the total acquisition duration for each echo between the GRE and EPI protocols and the limit of the gradient strength. Compared to GRE with shorter individual echo length or higher readout bandwidth, the current GRE protocol should perform similarly in the test-retest reliability. This is because the overall SNR and the sensitivity to physiological noise and instrumental instability is determined by the total acquisition time and echo time range rather than individual echo duration.

Compared to GRE with matched resolution and readout gradient strength, EPI with the longer echo train length is affected by more severe geometric distortion and blurriness due to $B_0$ inhomogeneity and $T_2^*$ signal decay, respectively. The EPI echo train length can be controlled by the EPI factor. Here, we discuss the scale of these effects. At 3 T, the maximum off resonant frequency of about 150 Hz in the brain was observed and resulted in 1.5 pixel or 1.5 mm distortion using the current 10 ms ETL as shown in Table 1. The effect of distortion correction as implemented in the Method section was evaluated in a simulation in the aspect of correction accuracy and algorithm-introduced blurring. In the simulation, the algorithm was used to correct for spatial shift of the discrete Delta function with various off-resonant $B_0$ values. The linear phase trend in the Fourier domain of the corrected "map" was used to quantify the correction accuracy, and the spectrum was used to define the effective resolution corresponding to the cutoff frequency at -3 dB. As shown in Fig. 8, the effective voxel size was shown to vary from 1 to 2 times the nominal voxel size depending on the actual distortion; assuming accurate field map, the corrected spatial shift agreed well with the prediction. In addition, blurring caused by the $T_2^*$ decay can be estimated according to the $T_2^*$ value and ETL (Qin, 2012). For example, with $T_2^*$ of 10 ms, which is shorter than the majority of the brain tissue even at 7 T, the full width at half maximum of the point spread function was estimated to be less than 1.2 voxels, representing limited blurring effect.

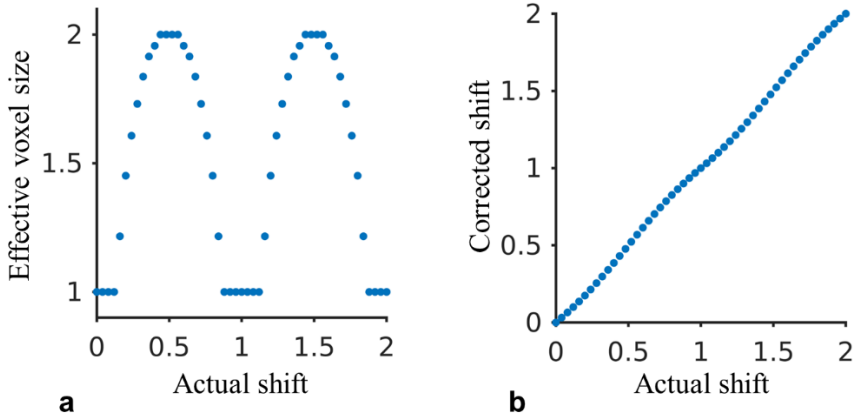

**Fig. 8** Evaluation of the effective voxel size (a) and accuracy (b) of the utilized distortion correction algorithm. All axes were normalized relative to the nominal voxels size.

The current study has several limitations. First of all, the protocols were only evaluated at one magnetic field strength. It did not consider the field-dependent physiological noise. Therefore, the quantitative result cannot be generalized to other field strength although the EPI method is expected to improve the test-retest reliability in general. Secondly, multi-shot EPI has a few challenges when used at very high resolution. The opposite gradient polarity between adjacent k-space lines introduces N/2 ghost artifact. Although it was corrected using phase data in the readout direction derived from the central k-space line data, this approach may not be sufficient when the effect in the non-readout directions is significant using strong readout gradient at higher resolutions. Additionally, the phase discontinuity along the edges of the k-space lines can also become more severe when the readout duration is long at high resolution. This can introduce ghost artifact in regions with significant off resonance or at higher field, which was shown to be reduced by averaging complex data from two scans with opposite readout gradient polarity (Stirnberg et al., 2022). Thirdly, the $R_2^*$ fitting was only based on the magnitude image. This was justified by sufficient SNR even for the last echo (>10) in the current data. It can introduce underestimation of $R_2^*$ if the SNR is much lower. Fourthly, the maximum achievable EPI factor and the associated test-retest reliability performance were not evaluated in this study. With gradient strength of 40 mT/m, slew rate of 200 mT/m/ms and ETL of 10 ms, EPI factor of 10 can be used. This is expected to further increase the test-retest reliability, but it may also lead to high gradient vibration, acoustic noise, peripheral nerve stimulation and poor patient comfort. As the gradient system develops (Beckett et al., 2022), this aspect remains to be evaluated.

**Conclusion**

Multi-shot multi-echo 3D EPI can be a useful alternative acquisition method for $T_2^*$w MRI and quantification of $R_2^*$ and $\chi$ with reduced scan time, improved test-retest reliability and similar accuracy compared to commonly used 3D GRE. In addition, the much faster EPI method is promising to pave the way for higher resolution $T_2^*$w MRI in clinical applications.

**Data Availability Statement**

In compliance with the requirements of the funding institute and with institutional ethics approval, the data and code can be made available upon reasonable request.

**Acknowledgements**

This work was partly supported by the Hamon Foundation and Texas Instrument Foundation and NIH/NIBIB (P41EB031771). The authors were grateful to Drs. Peter van Gelderen, Jacco de Zwart and Jeff Duyn in the advanced MRI section at the National Institute of Neurological Disorders and Stroke for sharing the MRI sequence and insightful discussion.